\documentclass[aps,prl,twocolumn,superscriptaddress]{revtex4-1}
\usepackage{graphicx}
\usepackage{amsmath}
\usepackage{amsfonts}
\usepackage{amssymb}
\usepackage{epsfig}
\usepackage{color}
\usepackage{bm}
\usepackage{wasysym}
\graphicspath{{pictures/}}
\begin{document}
	
\title{Ultrafast target charging due to polarization triggered by laser-accelerated electrons}
\author{A. V. Brantov}
\affiliation{P. N. Lebedev Physics Institute, Russian Academy of
Science, Leninskii Prospect 53, Moscow 119991, Russia}
\affiliation{Center for Fundamental and Applied Research,
Dukhov Research Institute of Automatics (VNIIA), Moscow 127055, Russia}
\author{A. S. Kuratov}
\affiliation{P. N. Lebedev Physics Institute, Russian Academy of
Science, Leninskii Prospect 53, Moscow 119991, Russia}
\affiliation{Center for Fundamental and Applied Research,
Dukhov Research Institute of Automatics (VNIIA), Moscow 127055, Russia}
\author{Yu. M. Aliev}
\affiliation{P. N. Lebedev Physics Institute, Russian Academy of
Science, Leninskii Prospect 53, Moscow 119991, Russia}
\author{V. Yu. Bychenkov}
\affiliation{P. N. Lebedev Physics Institute, Russian Academy of
Science, Leninskii Prospect 53, Moscow 119991, Russia}
\affiliation{Center for Fundamental and Applied Research,
Dukhov Research Institute of Automatics (VNIIA), Moscow 127055, Russia}

\begin{abstract}
A significant step has been made towards understanding the physics of the
transient surface current triggered by ejected electrons during the
interaction of a short intense laser pulse with a high-conductivity target.
Unlike the commonly discussed hypothesis of neutralization current
generation as a result of the fast loss of hot electrons to the vacuum, the
proposed mechanism is associated with excitation of the fast current by
electric polarization due to transition radiation triggered by ejected
electrons. We present a corresponding theoretical model and compare it with
two simulation models using the FDTD (finite-difference time-domain) and PIC
(particle-in-cell) methods. Distinctive features of the proposed theory are
clearly manifested in both of these models.
\end{abstract}
\maketitle

Strong ultra-fast surface fields and electric currents in the interaction of
short intense laser pulses with solid dense targets are currently of great
interest in both fundamental and applied science \cite{nakamura,mackena,Quinn,Sarri,Kar,Pompili}.
The generation of a strong surface current is closely related to lateral
electron transport, electromagnetic surface waves, and hot spot expansion on
the target. The last, for example, is important for ion acceleration. The
effect of lateral electron transport is considered in the context of the
physics of surface guided schemes for fast ignition \cite{nakamura}.
Excitation of electromagnetic surface waves is an elegant way to produce THz
radiation with well-concentrated energy \cite{Sakabe,Kuratov}, which is an
interesting complement to the widely discussed laser generation of THz
radiation into free space \cite{Hamster,Gopal}.

We especially emphasize the study with an entire series of experiments with
laser-triggered transient electric currents along a target surface
\cite{nakamura,mackena,Quinn,Sarri,Pompili}, to which our paper is closest.
Excitation of a lateral transient current has been standardly associated
with a loss of plasma neutrality because the laser-accelerated hot electrons
escape to the vacuum. This leads to generating a large positive potential,
which provides target neutralization via charge redistribution and the
antenna-like propagation of an electromagnetic disturbance away from
the interaction region \cite{Quinn}. The discharge of laser-irradiated
targets was thus assumed as a physical reason in several previous studies.
Unfortunately, no theoretical model has yet been developed to support such
a scenario. We do not intend to eliminate this shortcoming and instead aim
to highlight another mechanism for generating a lateral transient current,
which can be very effective.

Here, we present theoretical model based on the idea that a
laser-accelerated electron bunch crossing the target--vacuum boundary
generates a fast surface field and the corresponding lateral skin current
in the form of a polarization wave. Such a traveling-wave skin current
naturally appears in the same approach that was used in the theory of
transition radiation \cite{ginzburg} and is a fundamental effect for a
high-conductivity half-bounded medium, for example, a solid dense plasma.
We also complement our theory with two simulation models using FDTD
(finite-difference time-domain) and PIC (particle-in-cell) methods.
Distinctive features of the proposed theory are clearly manifested in both
of these models.

In our theoretical model, we assume that a bunch of laser-heated
relativistic electrons escapes the target with the velocity $v$ by crossing
the target--vacuum boundary and moves to infinity like those electrons from
a laser--plasma interaction that have sufficient energy to overcome a
blocking plasma potential. We describe this bunch as external normal to the
target surface current, ${\bf j}=en^e{\bf v}$, which appears at the target
surface at $t=0$, enters the vacuum, and detaches from a surface after some
time $\tau$. We assume that the target is characterized by the dielectric
permittivity $\epsilon$ and neglect all nonlinear effects related to the
target charging. This assumption is justified if the density $n^e$ of the
escaping electrons is much smaller than the target electron density $n_0$:
$n^e\ll n_0$. For example, this model can be applied to electrons leaving
the back of a thin target heated by short laser pulse (with the duration
$\tau_L$) and also to the front of a target with a preplasma whose thickness
does not prevent effective generation of electrons from a skin layer.

The full set of Maxwell equations for the generated electric field in the
Fourier space ($\omega,{\bf k}$) is reducible to the equations for the
normal ($\textbf{x}$) components $E_x^v$ ($x>0$) of this field in a vacuum
and $E_x^p$ ($x<0$) in a medium (irradiated target),
\begin{equation}\label{eq2}
\frac{\partial^2E_{x}^{p,v}}{\partial x^2}-k_{p,v}^2E_{x}^{p,v}=
\frac{4\pi e}{\epsilon}\frac{\partial\epsilon n^e_{\omega,k}}{\partial x}-
\frac{4\pi ei\omega v}{c^2}n^e_{\omega,k}\,,
\end{equation}
where ${\bf k}\perp\textbf{x}$, ${\bf v}=(v,0,0)$, $k_v=\sqrt{{\bf k}^2-
\omega^2/c^2}$, $k_p=\sqrt{{\bf k}^2-\epsilon \omega^2/c^2}$, and $\epsilon$
is the dielectric susceptibility. For $n^e(x-vt,{\bf r}_\perp)$, the
well-known solution \cite{ginzburg} of Eq.~\eqref{eq2}, which takes the
continuity of the tangential field components (both electric and  magnetic)
into account, $\omega$ as $\omega+i0$ and $n\equiv n(\omega,k)$ as given
by $n_{\omega,k}^e=ne^{i\omega x/v}/v$, can be written as
\begin{eqnarray} \nonumber \label{eqE}
E_{x}^v&=&-\frac{4\pi enkv}{v^2k_v^2+\omega^2}\left(\frac{i\omega}{kv}
\left(1-\frac{v^2}{c^2}\right)e^{i\frac{\omega x}{v}}+\right.
\\
&&\left.+\frac{k(\epsilon-1)(vk_p+i\omega(1-v^2/c^2))}{(k_p+\epsilon k_v)
(vk_p+i\omega)}e^{-k_vx}\right)\,,
\\\nonumber
E_x^p&=&-\frac{4\pi enkv}{\epsilon(v^2k_p^2+\omega^2)}
\left(\frac{i\omega}{kv}\left(1-\epsilon\frac{v^2}{c^2}\right)
e^{i\frac{\omega x}{v}}+\right.
\\\nonumber
&&\left.+\frac{k(\epsilon-1)(vk_v-i\omega(1-\epsilon v^2/c^2))}
{(k_p+\epsilon k_v)(vk_v-i\omega)}e^{k_px}\right)\,.
\end{eqnarray}
Solution \eqref{eqE} is a sum of the bunch field and the induced field (the
respective first and second terms). From the standpoint of physical effects,
Eq.~\eqref{eqE} describes the generation of transition radiation into free
space \cite{ginzburg_1946} and transition radiation along a surface \cite{eidman}.
Both types of radiation are well studied in the far-field approximation, but
the transient electromagnetic fields and electrical current have not yet
been studied, because the corresponding theory should be based on the
near-field theory. The development of such a theory is our main goal.
Experiments both on laser-triggered propagating electromagnetic pulses and
on electron transport along the conducting target \cite{nakamura,mackena,Quinn,Sarri,Kar,Pompili}
are urgently requested. Below, we use two-dimensional geometry in the space
$(x,z)$ to analyze this problem.

Generation of an electromagnetic surface wave is defined by the pole in
Eq.~\eqref{eqE}, i.e., the dispersion relation $d_0=k_p+\epsilon k_v=0$,
which has the solution $k_0\simeq\pm\omega/c(1-1/(2\epsilon))$ for
$|\epsilon|\gg1$. This contribution gives the expression for the normal
component of the vacuum-side surface wave field:
\begin{equation}\label{eqSW}
E_x^{sv}=-\int\frac{d\omega\,2ienv\omega}{c^2\sqrt{-\epsilon}}
e^{-i\omega(t\pm\frac{z}{c})}e^{-i\frac{\omega}{2\epsilon c}|z|}
e^{-\frac{\omega x}{c\sqrt{-\epsilon}}}\,.
\end{equation}
We note that surface field amplitude \eqref{eqSW} contains a small factor
$1/\sqrt{|\epsilon|}$ and hence disappears as $|\epsilon|\to\infty$.

\begin{figure}[!ht]
\centering{\includegraphics[width=8cm]{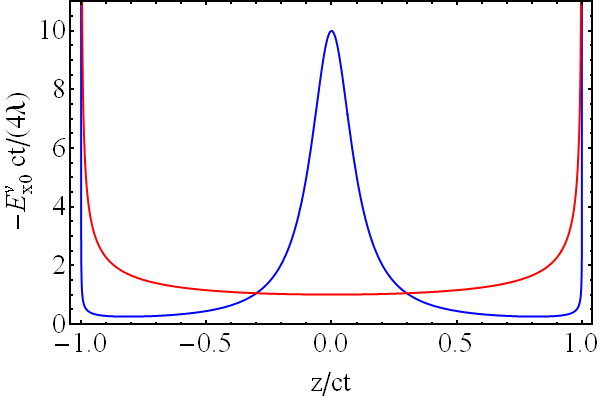}}
\caption{Vacuum-side surface electric field $E_{x0}^v(t,z)$ for $v=0.1c$
(blue line) and $v=0.99c$ ($\gamma\simeq7$) (red line).}
\label{fig1}
\end{figure}

In the limit of an ideal conductor $|\epsilon|\to\infty$ ($E_x^p=0$), the
normal component $E_{x0}^v= E_x^v(x=0)$ of the vacuum-side field at the
surface can be written as
\begin{equation}\label{eqx=0}
E_{x0}^v=E_r+E_d=-4\pi en\left(\frac{v}{c^2k_v}+
\frac{1-v^2/c^2}{vk_v-i\omega}\right)\,,
\end{equation}
where the term $E_r$ is responsible for the transition radiation field and
the term $E_d$ represents the dipole-like field of the uniformly moving
charge and its mirror image with respect to the surface \cite{bolotovskii}.
For a moving point charge, the spatial Fourier component of $E_r$ is
$E_r(k)=4\pi envJ_0(ckt)/c$, where $J_0$ is a Bessel function, which agrees
with the result in \cite{bolotovskii}. In the space $(x,z)$,
Eq.~\eqref{eqx=0} becomes
\begin{equation}\label{eqEs0}
E_{x0}^{v}=\begin{cases}-\displaystyle{\frac{4\lambda}{\sqrt{c^2 t^2-z^2}}}
\frac{cvt^2\gamma^2}{z^2+v^2t^2\gamma^2},&z^2<c^2t^2,\\
0,&z^2>c^2t^2,\end{cases}
\end{equation}
where $\lambda$ is the linear charge density and $\gamma=1/\sqrt{1-v^2/c^2}$
is the Lorentz factor of a moving (along $x$) filiform (along $y$) charge.
The polarization surface field propagates in the form of a transient wave at
the speed of light and has a sharp front at $z=ct$. This is shown in
Fig.~\ref{fig1}. The integrable divergence at the wave front in
Eq.~\eqref{eqEs0} is due to the singularity of the linear point-sized charge.

As it should be in the nonrelativistic limit $v\ll c$, polarization field
\eqref{eqEs0} is like a quasistationary dipole field of two charges (the
real charge and its mirror image) at the target surface under the trail of
the moving charge with two singular spikes from radiation field running in
opposite directions with negligible energy. In the ultrarelativistic limit,
this picture changes dramatically. For $v\backsimeq c$, the radiation field
$E_{x0}^{v}\backsimeq E_{r}\sim\lambda/\sqrt{c^2t^2-z^2}$ dominates the
dipole field (compare the red and blue lines in Fig.~\ref{fig1}) and spreads
away over the surface under the trail of the moving charge. We note that the
full surface charge $\lambda_s=(\int E_{x0}^v\,dz)/4\pi$ exactly equals the
escaping charge $\lambda$, i.e., $\lambda_s=-\lambda$. Bearing in mind the
laser generation of electrons, we below consider the ultrarelativistic case
$\gamma\gg1$.

The case of a point-sized source well highlights the difference in target
surface charging by slow and fast escaping electrons, which is directly
illustrated by the evolution of the normal component of the electric field
(Fig.~\ref{fig1}). Because the electric field $E_{x0}^v$ is exactly
proportional to the surface charge, it can be clearly seen that the
relativistic outgoing electron charge over time forms a positive surface
charging in the form of single waves moving apart, in contrast to the charge
spot spreading smoothly in the nonrelativistic case. We now show that the
same picture basically holds for a finite-sized electron charge generated by
a laser in both the normal (pulse width) and transverse (spot size)
directions. Assuming that $|\epsilon|,\gamma\gg1$ in Eq.~\ref{eqE} and using
the relations $B_y=(\omega\epsilon E_x+4\pi ij_x)/(k c)$ and
$E_z=ic/(\omega\epsilon)\partial B/\partial x$, we obtain the approximation
\begin{eqnarray}\label{eqE1}
\nonumber
&&E_{x}^v=-\frac{4\pi ene^{-k_vx}}{ck_v}\;,E_{x}^p=\frac{4\pi en}{\epsilon}
\!\left[\frac{e^{i\frac{\omega x}{c}}\!\!-\!e^{k_px}}{i\omega}\!-\!
\frac{e^{k_px}}{ck_v}\!\!\right ]\,,
\\
\nonumber
&&B_{y}^v=-\frac{4\pi en}{ck}\left(\frac{\omega}{ck_v}
e^{-k_vx}-ie^{i\frac{\omega x}{c}}\right)\,,
\\
&&B_{y}^p=-\frac{4\pi en}{ck}\left(\frac{\omega}{ck_v}
e^{k_px}-ie^{k_px}\right)\,,
\\
\nonumber
&&E_{z}^v=\frac{4\pi ien}{ck}\left(e^{-k_vx}\left[1+\frac{\omega}{ck_p}
\left(\frac{\omega}{ck_v}-i\right)\right]-e^{i\frac{\omega x}{c}}\right)\,,
\\
\nonumber
&& E_{z}^p=\frac{4\pi i\omega en}{c^2kk_p}
\left(\frac{\omega}{ck_v}-i\right)e^{k_p x}\,.
\end{eqnarray}

\begin{figure}[!ht]
\centering{\includegraphics[width=8cm]{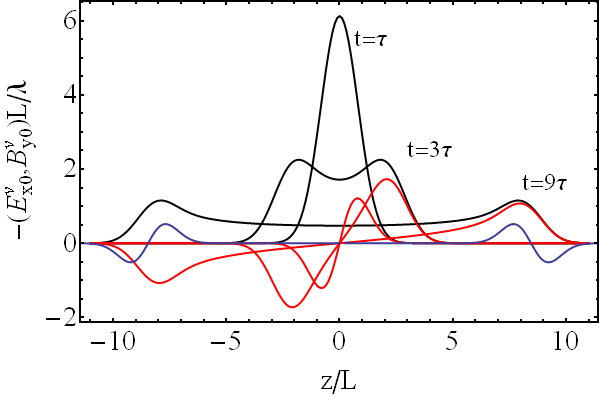}}
\caption{Evolution of the vacuum electric field $E_{x0}^v(t,z)$ (black
curves) and magnetic field $B_{y0}^v(t,z)$ (red curves) at the
target--vacuum boundary for $L=c\tau$: the blue curve shows the electric
field of the electromagnetic surface wave (see Eq.~\ref{eqSW}) for
$\omega_{p}\tau=10$ at $t=9 \tau$.}
\label{fig2}
\end{figure}
To illustrate the generation and evolution of a polarization wave by a
finite-sized laser-accelerated electron bunch, we choose the plasma
permittivity $\epsilon=1\!\!-\!\omega_{p}^2/\omega^2\!\backsimeq\!\!-
\omega_{p}^2/\omega^2$ and the current source $e n^e=\lambda\theta(t)
(\theta(vt-x)-\theta(vt-c\tau-x))\exp(-z^2/L^2)/(Lv\tau\sqrt{\pi})$. Here,
the bunch with $v\simeq c$ enters the vacuum at $t=0$ and leaves the target
at $t=\tau$, and the Heaviside step functions $\theta(\xi)$ model the
laser-heated spot size $L$ and pulse duration $\tau$. Using $en=i\lambda
e^{-k^2L^2/4}(1-e^{i\omega\tau})/(\omega\tau)$ in Eqs.~\eqref{eqE1}, we
studied the evolution of the polarization wave, which is shown in
Fig.~\ref{fig2}. The amplitude of the induced electric field at the surface
initially ($t<\tau$) increases inside the interaction spot to the maximum
value defined by the ratio $L/c\tau$ and at $L=c\tau$ is $\sim 6\lambda/L$.
After a pulse terminates ($t>\tau$), the field separates into two wave
bunches propagating at the speed of light in opposite directions along the
target surface away from the interaction spot. During propagation, the field
amplitude decreases with time as $1/t$. In accordance with Fig.~\ref{fig2},
an electron charge of $\sim100pC/\mu m$ escaping from a spot with
$L\sim5\mu m$ can generate a surface electric field up to TV/m.

\begin{figure}[!ht]
\centering{\includegraphics[width=8cm]{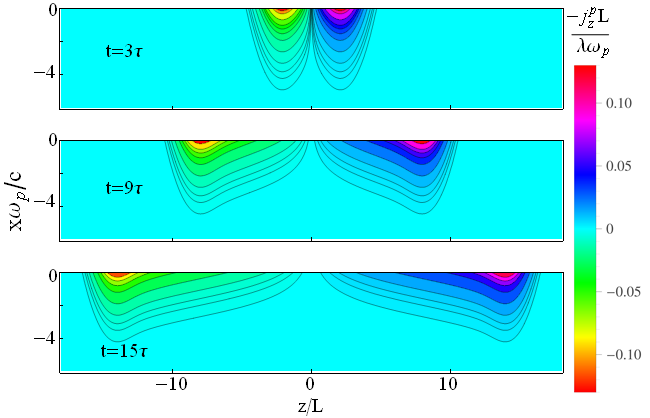}}
\caption{Space--time distribution of the electric current density
$j_z^pL/(\lambda\omega_p)$ for $L=c\tau$.}
\label{fig3}
\end{figure}

The tangential field $E_z^p\propto1/\sqrt{|\epsilon|}$ corresponds to the
strong plasma current $j_z^p\backsimeq-i\omega\epsilon E_z^p/(4\pi)\propto
\sqrt{|\epsilon|}$ associated with the polarization wave, which behaves
similarly to the magnetic field $B_y^p$ in Eq.~\eqref{eqE1}. For $t>\tau$,
we can write the electron current as
\begin{equation}\label{eqjz}
j_z^p\!=\!\frac{i\lambda\omega_p}{2c\tau}\!\!\int\!\!
\frac{dk}{\pi k}[J_0(ck(t\!-\!\tau))\!-\!J_0(ckt)]
e^{-\frac{k^2L^2}{4}}e^{ikz+\frac{cx}{\omega_p}}.
\end{equation}
This current runs at the speed of light inside the skin layer in the form of
two unipolar pulses propagating away from the interaction spot as shown in
Fig.~\ref{fig3}. The full linear charge inside these two pulses is exactly
equal to the linear charge $\lambda$ of escaping electrons. Because the
polarization-triggered current is excited in a neutral plasma, we interpret
it as a charging current unlike the previously assumed charge-neutralizing
current \cite{Quinn}. The polarization wave field is a high-frequency field,
whose characteristic frequency $\omega_*$ can be estimated as
$\omega_*\simeq\min\{\tau^{-1},c/L\}$.

Because the polarization wave amplitude decreases monotonically $\propto1/t$,
it eventually drops below the surface wave amplitude, which was initially
smaller by the factor $1/\sqrt{|\epsilon|}$. These amplitudes become
comparable at the instant $t_s\sim \sqrt{|\epsilon|}/\omega_*$, where
$|\epsilon|$ should be evaluated at the frequency $\sim\omega_*$ or,
equivalently, at the distance $l_s\sim ct_s$. For a plasma target, $t_s\sim
\omega_{p}/\omega_*^2$, and for the metal target, $t_s\sim\omega_*^{-3/2}
\sqrt{\sigma}$, where $\sigma$ is the electrical conductivity at $\omega\sim
\omega_*$.  The distance $l_s$ increases rather quickly with the laser pulse
duration $\tau$ if $L/c<\tau$, and for a 1\,ps pulse reaches a few cm to
several tens of cm. At a large distance $z\gtrsim l_s$, the transient
polarization wave disappears, and only the surface wave remains. For a
plasma target, the latter is described by the analytic expression
\begin{equation}\label{eqSW1}
E_x^{sv}=\frac{4\sqrt{\pi}\lambda}{\omega_p\tau L}
\left(e^{-\frac{(ct\mp z)^2}{L^2}}-
e^{-\frac{(ct-c\tau\mp z)^2}{L^2}}\right)\theta(\pm z)\,.
\end{equation}
We note that in deriving Eq.~(\ref{eqSW1}), we neglect the imaginary part of
$\epsilon$. A more accurate calculation shows that the surface wave
propagates with a velocity slightly different from the speed of light in the
form of a bipolar pulse (a two-pi pulse) unlike the polarization wave as
seen in Fig.~\ref{fig2}.

\begin{figure}[!ht]
\centering{\includegraphics[width=8cm]{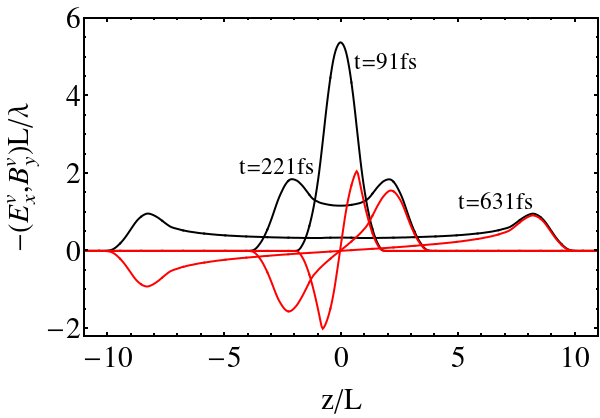}}
\caption{Evolution of the vacuum electric field $E_{x0}^v(t,z)$ (black
curves) and magnetic field $B_{y0}^v(t,z)$ (red curves) at the
target--vacuum boundary from FDTD modeling for $L=c\tau=20\,\mu$m.}
\label{fig4}
\end{figure}

Our theory ignores the possible permanent loss of a small fraction of the
laser-accelerated electrons from a target hot spot. The theoretical
description of the related charging surface field and current still remains
unclear, although such a postulation has been attributed to the experiment
\cite{Quinn}. To clarify whether a charge-neutralizing current could be
detectable, we performed three-dimensional FDTD numerical simulations using
the code VSim. To advance the study of a transient field propagating along
the target at the speed of light, we applied this simulation to a metal
target, where the dielectric permittivity is a complex function
($\epsilon\prime+i\epsilon\prime\prime$) but still $|\epsilon|\gg1$. We
adopted the same model for the escaping electron beam uniform along the $y$
axis as above with $v=0.99c$, $L=c\tau=20\,\mu$m, and $q/L=100$\,pC/$\mu$m
for $\epsilon$ given by the standard Drude model: $\epsilon=1+4\pi i\sigma
(\omega)/\omega$, where $\sigma=\sigma_0/(1-i\omega/\nu)$ with $\sigma_0=
10^{18}$\,s$^{-1}$ and $\nu=10^{13}$\,s$^{-1}$. The target occupied a
half-space $x<75\,\mu$m in the full simulation box $0<x<300\,\mu$m,
$-300\,\mu$m$<y<300\,\mu$m, and $-300\,\mu$m$<z<300\,\mu$m. The grid cell
size was $1\,\mu$m and the time step was 1\,fs.

In contrast to the analytic model, we assumed full charge neutrality in the
FDTD simulation resulting in target charging during electron escape. We show
our simulation results in Fig.~\ref{fig4}, which agree well with the
analytic model showing that the polarization wave is the main contributor to
the surface electromagnetic field.

\begin{figure}[!ht]
\centering{\includegraphics[width=4.1cm]{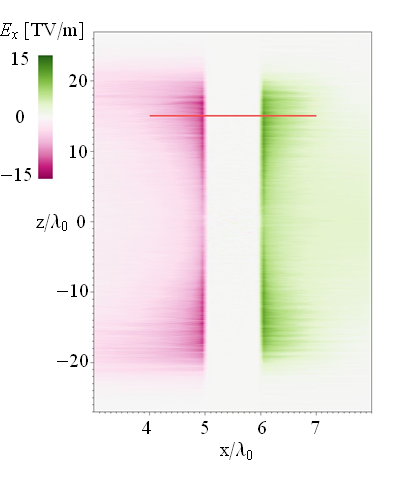} \includegraphics[width=4.1cm]{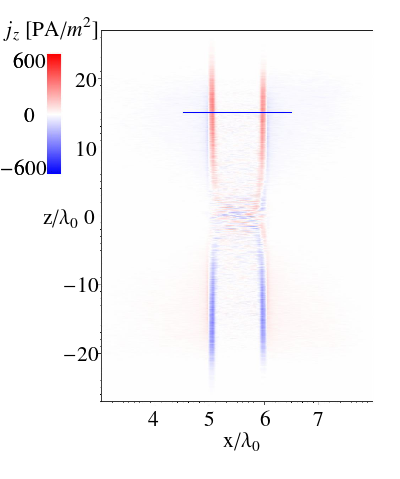} \includegraphics[width=7.cm]{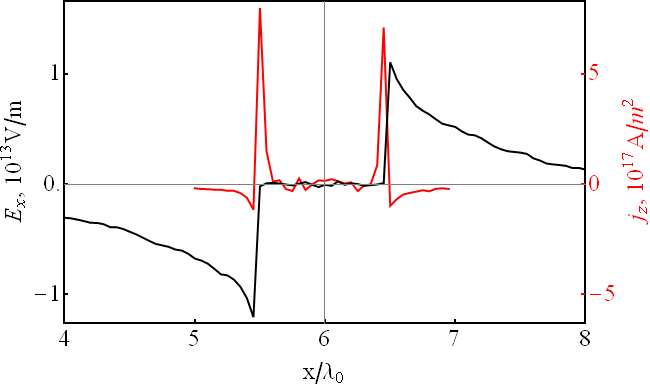} }
\caption{Spatial distributions (from PIC simulation) of the electric field
(left) and current (right) in the upper panel together with their
cross-section profiles (along the shown line) in the bottom panel at
t=190\,fs.}
\label{fig5}
\end{figure}

In accordance with the analytic theory, this simulation shows that a change
in the type of dielectric permittivity has only a minor effect on the
surface field distributions until $|\epsilon|\gg1$. In a similar simulation
where we used an escaping bunch with a small velocity $0.2c$, we observed a
barely smooth spreading of the field rather than a wavelike structure, which
also corresponds to the analytic theory.

We also performed a two-dimensional PIC simulation of the interaction of a
short laser pulse with a plasma slab target. The laser pulse (Gaussian in
both space and time with a 30\,fs FWHM duration, a $10^{21}$\,W/cm$^2$
maximum intensity, and the wavelength $\lambda_0=1\,\mu$m) was focused on the
FWHM spot $L=3\lambda_0$ at the target front side ($x=5\lambda_0$, $z=0$ in
Fig.~\ref{fig5}). We used a target thickness of $1\lambda_0$ and an electron
density of $100n_c$, where $n_c$ is the critical density, immovable ions,
and 10 macroparticles per cell for electrons. The full simulation box was
$8\lambda_0\times100\lambda_0$ with a spatial resolution of $\lambda_0/200$
in both directions. The PIC simulation results also confirm formation of a
surface polarization wave propagating along both sides of the target in
opposite directions, as shown in Fig.~\ref{fig5}. The maximum surface
electric field initially reaches $\sim15$\,TV/m in agreement with our
theoretical model for $\lambda\sim2$\,nC/$\mu$m shown by the PIC result.
This field decreases similarly to $\propto1/z$ and predictably drops to the
multi-GV/m level at  distance of $\sim$1cm. In this regard, we note that
electromagnetic pulses propagating at the speed of light with a strong
electric field (of the order of GV/m) have already been measured at a cm
distance from the hot spot \cite{Kar}. As a final note, we emphasize that a
small fraction of the laser-accelerated hot electron population expands
along the target surface from the vacuum side because it is well held by the
surface fields (see \cite{mackena,Quinn,nakajima13}).

In summary, our theory sheds light on the physical mechanism of the
generation and propagation of a transient electromagnetic pulse and a
lateral current in the wave form along the target surface at the speed of
light. The proposed mechanism is associated with fast electric polarization
of a high-conductivity target during ejection of a laser-driven electron
bunch from a target into a vacuum. Supported by the results of the developed
theory and two simulation models, the mechanism proposed and studied here
might be important for a deeper understanding of the experiments, which have
already shown the extremely strong ultrafast charging of a solid irradiated
by a high-intensity laser and the propagation of the corresponding transient
laser-triggered current along the target surface \cite{mackena,Quinn,Sarri,Kar,Pompili,Sakabe}.

This research was supported by the Russian Science Foundation.

\end{document}